\documentclass[%
 reprint,
nofootinbib,
nobibnotes,
bibnotes,
 amsmath,amssymb,
 aps,
prl,
superscriptaddress,
]{revtex4-1}

\usepackage{graphicx}
\usepackage{dcolumn}
\usepackage{bm}
\usepackage{hyperref}

\def\mnras{Mon. Not. R. Astron. Soc.}%
\def\apj{Astrophys. J.}
\def\apjl{Astrophys. J. Lett.}
\def\prl{Phys. Rev. Lett.}

\begin{document}

\title{Revealing the formation of stellar-mass black hole binaries:\\The need
for deci-Hertz gravitational wave observatories}

\author{Xian Chen}
\email{xian.chen@pku.edu.cn}
\affiliation{Astronomy Department, School of Physics, Peking University, 100871 Beijing, China}
\affiliation{Kavli Institute for Astronomy and Astrophysics at Peking University, 100871 Beijing, China}
\author{Pau Amaro-Seoane}
\email{pau@ice.cat}
\affiliation{Institut de Ci{\`e}ncies de l'Espai (CSIC-IEEC) at Campus UAB, Carrer de Can Magrans s/n 08193 Barcelona, Spain\\
Kavli Institute for Astronomy and Astrophysics at Peking University, 100871 Beijing, China\\
Institute of Applied Mathematics, Academy of Mathematics and Systems Science, CAS, Beijing 100190, China\\
Zentrum f{\"u}r Astronomie und Astrophysik, TU Berlin, Hardenbergstra{\ss}e 36, 10623 Berlin, Germany
}

\date{\today}

\begin{abstract}
The formation of compact stellar-mass binaries is a difficult, but interesting
problem in astrophysics.  There are two main formation channels: In the field
via binary star evolution, or in dense stellar systems via dynamical
interactions.  The Laser Interferometer Gravitational-Wave Observatory (LIGO)
has detected black hole binaries (BHBs) via their gravitational radiation.
These detections provide us with information about the physical parameters of
the system. It has been claimed that when the Laser Interferometer Space
Antenna (LISA) is operating, the joint observation of these binaries with LIGO
will allow us to derive the channels that lead to their formation.  However, we
show that  for BHBs in dense stellar systems dynamical interactions could
lead to high eccentricities such that a fraction of the relativistic mergers
are not audible to LISA.  A non-detection by LISA puts a lower limit of about
$0.005$ on the eccentricity of a BHB entering the LIGO band. On the other
hand, a deci-Hertz observatory, like DECIGO or Tian Qin, would significantly
enhance the chances of a joint detection, and shed light on the formation
channels of these binaries. 
\end{abstract}

\maketitle

{\it Introduction.}--The first LIGO events, GW150914 and GW151226
\citep{ligo16a,ligo16b}, are consistent with mergers of General-Relativity
black holes (BHs). Data analysis reveal that the orbits
started at a semi-major axis of $a\sim10$ Schwarzschild radii ($R_S$) with an
eccentricity of $e<0.1$. The BH masses are about $M_1\simeq36$ and
$M_2\simeq29~M_\odot$ for GW150914 and $M_1\simeq14$ and $M_2\simeq7.5~M_\odot$
for GW151226.  The detections can be used to infer new, more realistic event
rates, of about $9-240~{\rm Gpc^{-3}~yr^{-1}}$ \citep{ligo16rate}. This rate
agrees with two formation channels: (i) evolution of a binary of two stars in
the field of the host galaxy, where stellar densities are very low (e.g
\cite{belczynski16}) or (ii) via exchange of energy and angular momentum in
dense stellar systems, where the densities are high enough for stellar close
encounters to be common (e.g. \cite{rodriguez16}).

LIGO and other ground-based gravitational wave (GW) observatories, such as
Virgo, are, however, blind with regarding the formation channels of BH binaries
(BHBs). Both channels predict populations in the $10-10^3~{\rm Hz}$ detector
band with similar features, i.e. masses larger than the nominal
$10\,M_{\odot}$, a mass ratio ($q\equiv M_2/M_1$) of about $1$, low spin, and
nearly circular orbits \cite{ligo16astro,ama16}.

It has been suggested that a joint detection with a space-borne observatory
such as LISA
\cite{Amaro-SeoaneEtAl2012,Amaro-SeoaneEtAl2013,Amaro-SeoaneEtAl2017} could
allow us to study different moments in the evolution of BHBs on their way to
coalescence:  LISA can detect BHBs when the BHs are still $10^2-10^3~R_S$
apart, years to weeks before they enter the LIGO/Virgo band
\cite{miller02,ama10,kocsis12levin,kocsis13,sesana16,seto16,vitale16}.  At such
a separation, the orbital eccentricity bears the imprint of the formation
channel because (i) BHBs in dense stellar systems form on systematically more
eccentric orbits and (ii) the GW radiation at this stage is too weak to
circularize the orbits \cite{miller02,wen03,gultekin04,gultekin06,oleary06}.
Therefore, circular binaries typically form in the field, while eccentric ones
through the dynamical channel.  Recent studies further predict that those BHBs
with an eccentricity of $e>0.01$ in the LISA band preferentially originate from
the dynamical channel
\cite{kyutoku16,nishizawa16a,nishizawa16b,breivik16,seto16}.

In this letter we prove that eccentric BHBs originating in dense stellar
environments have a large chance to elude the LISA band.

\begin{figure}
\includegraphics[width=1\columnwidth]{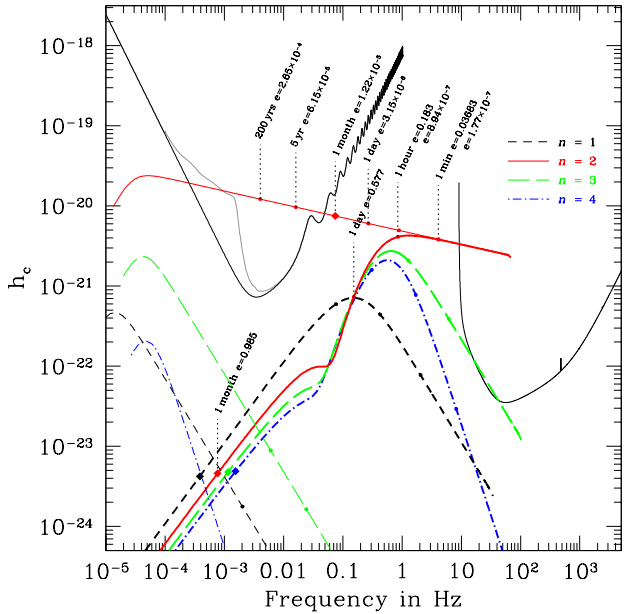}
\caption
   {
Characteristic amplitude $h_c$ of the first four harmonics (indicated with
numbers) emitted by a BHB with masses $M_1=M_2=30\,M_{\odot}$ and at a
luminosity distance of $D=500~{\rm Mpc}$.  The amplitude is calculated as
described in \cite{barack04} and the orbital evolution as in \cite{peters64}.
We display a BHB starting at a semi-major axis of $a_0=0.1$ AU and with initially two
very different eccentricities, so as to illustrate the main idea of this article:
(i) $e_0=0.05$ (thin colored lines), and (ii) an extreme case, $e_0=0.999$ (thick
colored lines).  Along the harmonics we mark several particular moments with
dots, where the labels show the time before the coalescence of the binary and
the corresponding orbital  eccentricities.  The two black solid curves depict
the noise curves ($\sqrt{f\,S_h(f)}$) for LISA and LIGO in its advanced
configuration. Although we have chosen a very high eccentricity for the second
case in this example,
we note that lower eccentricities can also be inaudible to LISA (see discussion).
\label{fig.harmonics}} \end{figure}

\textit{Inaudible black hole binaries}--Non-circular BHBs have two distinct
properties. (i) Eccentricity damps the characteristic amplitude ($h_c$) of each
GW harmonic, as compared to a circular BHB. In Figure~\ref{fig.harmonics} we
depict two sources similar to GW150914 but originating from two distinct
channels, i.e. with two different initial eccentricities. In the
low-eccentricity case, the $n=2$ harmonic predominates and it is strong enough
to be jointly detected by LISA and LIGO/Virgo.  In the (very) eccentric case,
however, the amplitudes of the harmonics are orders of magnitude below the
noise level of LISA, so that a joint detection is ruled out. When the
eccentricity has been significantly damped, about one hour before the merger,
the dominant harmonic starts to converge to the $n=2$ one, and later, upon
entering the LIGO band, becomes indistinguishable from that in the circular
case.
Therefore, the imprint about the formation channel is lost.

(ii) Increasing the eccentricity shifts the peak of the relative power of the
GW harmonics towards higher frequencies (see Fig. 3 of \cite{Peters63}).
Hence, more eccentric orbits emit their maximum power at frequencies farther
away from LISA. More precisely, when $e=0$, all the GW power is
radiated through the $n=2$ harmonic, so that the GWs have a single
frequency of $2/P$, where $P=2\pi(GM_{12}/a^3)^{-1/2}$ is the orbital period
and $M_{12}=M_1+M_2$.  On the other hand, when $e\simeq1$, the
$n=2.16(1-e)^{-3/2}$ harmonic becomes predominant \cite{farmer03}, so most GW
power is radiated at a frequency of $f_{\rm peak}=2.16(1-e)^{-3/2}P^{-1}$.

\begin{figure}
\includegraphics[width=1\columnwidth]{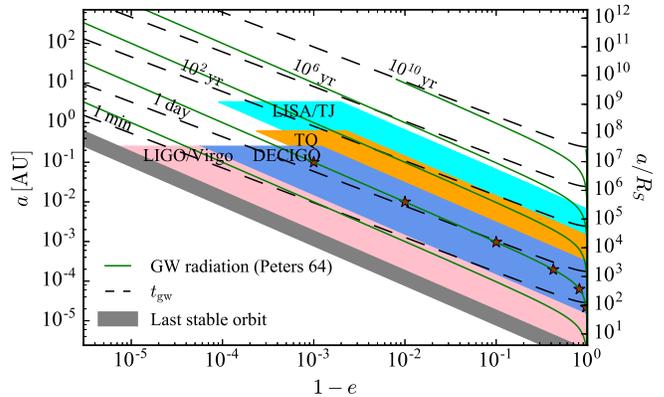} \caption{ Different
detectors' bands for a binary of $M_1=M_2=30~M_\odot$. We have considered four
types of detectors: (i) a ground-based interferometer like LIGO and Virgo (pink
stripe), with the minimum and maximum observable frequencies
$(f_1,\,f_2)\sim(10,10^3)~{\rm Hz}$ \cite{abbott09,accadia10}, (ii) a
space-borne solar-orbit interferometer such as the DECi-hertz Interferometer
Gravitational Wave Observatory (DECIGO, blue) with $(f_1,f_2)\sim(0.1,10)~{\rm
Hz}$ \cite{kawamura11}, (iii) a geocentric space observatory like the Tian Qin
project (TQ hereafter, orange) with T$(f_1,f_2)\sim(10^{-2},0.3)~{\rm Hz}$
\citep{luo16}, and (iv) another solar-orbit interferometer but with
million-kilometer baseline, like LISA or Tai Ji (TJ hereafter, shown as cyan),
which operates at milli-Hz, $(f_1,f_2)\sim(10^{-3},0.1)~{\rm Hz}$
\citep{Amaro-SeoaneEtAl2013,gong15} The upper, horizontal limit in the color stripes
corresponds to an orbital period of one week for LIGO/Virgo/DECIGO, one month
for TQ, and one year for LISA/TJ, as imposed by the restrictions in the search
of the different data streams.  The green solid lines show the evolutionary
tracks of a binary evolving only due to GW emission, in the approximation of
Keplerian ellipses \cite{peters64}. The dashed, black lines are isochrones
displaying the time to relativistic merger in the same approximation ($t_{\rm gw}$, see text), provided
that the evolution is driven only by GWs. The thick gray stripe displays the
last stable orbit, below which the two BHs will merge within one orbital period.
We also display with red stars the positions of the eccentric BHB in
Figure~\ref{fig.harmonics}
at different stages, to illustrate the process.
  \label{fig:detectors}}
\end{figure}

In Figure~\ref{fig:detectors} we display the $a-(1-e)$ plane for a BHB.  The
boundaries of the stripes have been estimated by looking at the minimum and
maximum frequencies audible by the detectors, $f_1$ and $f_2$, and letting
$f_1<\,f_{\rm peak}<\,f_2$, with $f_{\rm peak}$ defined before.  If a BHB is
evolving only due to GW emission, it will evolve parallel to the green lines.
These  track are parallel to the stripes because as long as $e\simeq1$, the
pericenter distance, $r_p=a\,(1-e)$, is almost constant during the evolution
\cite{peters64}, and a constant $r_p$ corresponds to a constant $f_{\rm peak}$.
Because of this parallelism, a BHB cannot evolve into the band of a GW detector
if it initially lies below the detector stripe.

Hence, we can see that some binaries will fully miss the LISA/TJ range. A good
example is the eccentric BHB we chose for Figure~\ref{fig.harmonics}. A
detector operating at higher frequencies, such as TQ or DECIGO, can however
cover the relevant part of the phase-space, so that a joint search is possible.
These detectors could alert LIGO/Virgo decades to hours before an event is
triggered, as one can read from the isochrones of Figure~\ref{fig:detectors}.

{\it Dense stellar environments.--}BHBs such as the one we have used for our
last example completely miss the LISA/TJ band. Eccentric binaries typically
originate from dense stellar systems such as globular clusters (GCs) and
nuclear star clusters (NSC), as shown by a number of authors in a number of
publications
\citep{miller02,wen03,gultekin04,gultekin06,oleary06,nishizawa16a,nishizawa16b,breivik16}.
In these systems, BHs diffuse towards the center via a process called mass
segregation \cite[see
e.g.][]{Peebles72,BW76,ASEtAl04,FAK06a,AlexanderHopman09,PretoAmaroSeoane10}.
To model it, we adopt a Plummer model \cite{Plummer11}, and we assume that the
mean stellar density is $\rho_*=5\times10^{5}~M_\odot~{\rm pc^{-3}}$ and the
one-dimensional velocity dispersion is $\sigma_*=15~{\rm km~s^{-1}}$. These
values correspond to a typical GC with a final mass of $M_{\rm
GC}\approx10^5~M_\odot$ and a half-mass radius of $R_h\approx0.5$ pc.  We note,
however, that the main conclusions derived in this work do not significantly
change for a NSC.

The two driving and competing mechanisms in the evolution of any BHB in the
center of the cluster are (i) interaction with other stars, ``interlopers'',
which come in at a rate of $\Gamma\sim2\pi G\rho_*a(M_{12}/M_*)/\sigma_*$, with
$M_*=10~M_\odot$ the mean mass of the interlopers because the cluster has gone
through mass segregation, and (ii) gravitational radiation, which shrinks the
orbital semi-major axis at a rate of

\begin{equation}
\dot{a}_{\rm gw}=-\frac{8\,c\,R_S^3q\,(1+q)}{5a^3(1-e^2)^{7/2}}
\left(1+\frac{73}{24}e^2+\frac{37}{96}e^4\right),
\end{equation}

\begin{figure}
\includegraphics[width=1\columnwidth]{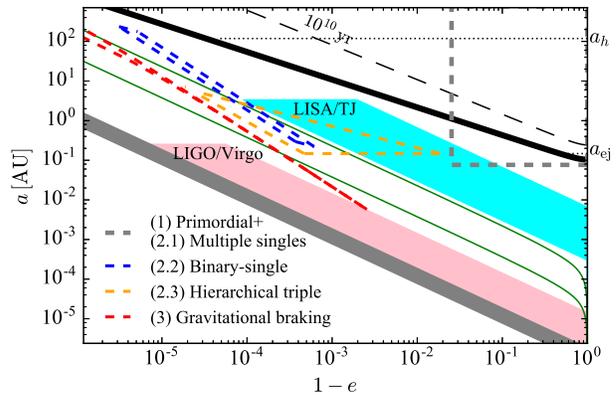}
\caption{
Phase space structure of a BHB with $M_1=M_2=30~M_\odot$. The top-right box
fences in the birthplace of 95\% of a thermal distribution of primordial
binaries, i.e. those binaries formed not dynamically but via binary stellar
evolution. In this box, but limited within the radii $a_h$ and $a_{\rm ej}$,
the hard and ejection radius, which end at the boundary of the dynamical region because of the absence of interlopers, we also find the vast majority of binaries formed
dynamically, i.e. the 95\% of their thermal distribution. The colored, dashed
lines depict the birthplaces of BHBs formed via three different processes which
we explain in the main text. The green lines display the evolutionary tracks of
a BHB entering the LIGO/Virgo band  at two different eccentricities, 
$e=0.1$ (lower) and 
$e=5\times10^{-3}$ (upper). The first LIGO detections have an
eccentricity $e\lesssim0.1$, meaning that they have formed
between the lower green line and the upper thick, black line.
\label{fig:GC}}
\end{figure}

\noindent
\cite{peters64}. We can readily separate the phase-space in two distinct
regimes according to these two competing processes by equating their associated
timescales: $t_{\rm int}:=1/\Gamma$ and $t_{\rm
gw}:=(1/4)\,\left|a/\dot{a}_{\rm gw}\right|$, which defines the threshold shown
as the thick, black line in Figure~\ref{fig:GC}. The reason for the $1/4$ factor
is given in \cite{peters64}.  Below the curve, BHBs will evolve due to GW
emission. Above it, close encounters with interlopers are the main driving
mechanism, so that BHBs can be scattered in
both directions in angular momentum in a random-walk fashion. The
scattering in energy is less significant but also present (see
\cite{alexander17} and discussion in \cite{Amaro-SeoaneLRR2012}).

{\it Possible ways of forming relativistic BHBs.--} Different mechanisms have
been proposed in the literature to form a BHB which eventually might end up
emitting detectable GWs.

\noindent
(1) Primordial binaries: In stellar dynamics this term refers to binaries
already present in the cluster which form via stellar evolution.  Population
synthesis models predict that these binaries populate the area of phase-space
displayed as the grey thick-dashed box of Figure~\ref{fig:GC} (see e.g.
\cite{belczynski04}). We note that only a small fraction of them are in the
LISA/TJ band.

\noindent
(2) Dynamics: (2.1) Close encounters of multiple single, i.e. initially not
bound, objects also form BHBs (see e.g.
\cite{kulkarni93,sigurdsson93,miller02hamilton,ivanova05}). Their formation
follows a thermal distribution in $e$ (e.g. \cite{antognini16}), like
primordial binaries, but the distribution of $a$ is better constrained: When
the binding energy of the binary, $E_b=GM_1M_2/(2a)$ becomes smaller than the
mean kinetic energy of the interlopers, $E_*=3M_*\sigma_*^2/2$, the binary
ionizes \cite{BT08}. The threshold condition $E_b=E_*$ can be expressed in
terms of a ``hard radius'', $a_h=GM_1M_2/(3M_*\sigma_*^2)$.  These ``hard''
binaries heat up the system, meaning that they deliver energy to the rest of
the stars interacting with them: Binaries with $a<a_h$ impart on average an energy
of $\Delta E\simeq kG\mu M_*/a$ to each interloper, where $\mu$ is the reduced
mass of the binary and $k$ is about $0.4$ when $M_1\simeq M_2\simeq M_*$
\citep{heggie75}.  The interloper hence is re-ejected into the stellar system
with a higher velocity because of the extra energy,
$v\sim\left(3\sigma_*^2+2kG\mu/a\right)^{1/2}$, and the center-of-mass of the
BHB recoils at a velocity of $v_b\sim M_*v/(M_1+M_2)$.  Occasionally, the BHB
will leave the system if this velocity exceeds the escape velocity of the GC,
$v_{\rm esc}=\sqrt{2.6GM_{\rm GC}/R_h}$ \cite{rodriguez16}. The threshold for
this to happen is defined by the condition $v_b=v_{\rm esc}$, i.e. the binary
must have a semi-major axis smaller than the ``ejection radius'', $a_{\rm ej}$.
Therefore, all of these BHBs are confined in $a_h<a<a_{\rm ej}$ of
Figure~\ref{fig:GC}. Because of their thermal distribution, we have that
$95\%$ of them have $e<0.975$. Therefore, they populate an even smaller area
than those primordial binaries.

\noindent
(2.2) Binary-single interactions: Initially we have a hard BHB which interacts
with a single object in a chaotic way. During the interaction the
interloper might excite the eccentricity of the inner binary to such high
values that the binary is on an almost head-on-collision orbit, to soon merge
and emit a detectable burst of GWs \cite{gultekin06,samsing14,ama16}. This
happens only if $t_{\rm gw}$ is shorter than the period of the captured
interloper $P_{\rm int}$. The event rate for BHBs has not been calculated for this
scenario but earlier calculations for neutron-star binaries find it to be
$1~{\rm Gpc^{-3}~yr^{-1}}$ \cite{samsing14}.  We derive now the
eccentricities of these BHBs: Suppose the
semi-major axis of a BHB changes from $a$ (with, of course, $a_{\rm
ej}<\,a<\,a_h$) to $a'$, and $e$ to $e'$ during the three-body interaction, and
the final orbit of the interloper around the center-of-mass of the BHB has a
semi-major axis of $a_{\rm int}$.  Energy conservation results in the following
relations,  $a'>a$ and $a_{\rm int}\simeq 2a/(1-a/a')$ (see \cite{samsing14}),
where we neglect the initial energy of the interloper because the BHB is
assumed to be hard.  Then using a conservative criterion for a successful
inspiral, $t_{\rm gw}(a',\,e')=P_{\rm int}(a_{\rm int})$, we derive $e'$ for the BHB,
which allows us to confine the range of eccentricities as the dashed, blue
curve of Figure~\ref{fig:GC}.

\noindent
(2.3) Hierarchical triple: This is similar to the previous configuration, but
now we only consider $1<a'/a<1.5$, because this requires that $a_{\rm int} >
6\,a$, in which case the  configuration is stable \cite{mardling01}.  This
leads to a secular evolution of the orbital eccentricity of the inner BHB which
is known as the Lidov-Kozai resonance (see \cite{lidov62,kozai62} and also
\cite{miller02,wen03,oleary06,naoz13,antognini14,liu15}).  The inner BHB will
decouple via GW emission and merge at a critical eccentricity, and the merger
rate has been estimated to be $0.3-6~{\rm Gpc^{-3}~yr^{-1}}$
\cite{antonini14,antonini16BS,kimpson16,silsbee16}.  We follow the scheme of
\cite{antonini14} of isolated hierarchical triples but impose four additional
requirements which are fundamental for a realistic estimation of the threshold
eccentricity in our work: (a) The BHB has $a_{\rm ej}<\,a<\,a_h$.  (b) The third body
orbiting the BHB has a mass of $M_{\rm int}=10~M_\odot$ because of mass
segregation, and an eccentricity of $e_{\rm int}=2/3$, which corresponds to the
mean of a thermal distribution \cite{antognini16}.  (c) The outer binary, i.e.
the third object and the inner BHB, is also hard, so that $a_{\rm
int}<GM_{12}/(3\sigma_*^2)$. (d) The pericenter distance of the outer binary,
$a_{\rm int}(1-\,e_{\rm int})$ should meet the criterion for a stable triple
(Eq. 90 in Ref.  \cite{mardling01}). These conditions delimit the range of
eccentricities as shown by the dashed, orange lines in Figure~\ref{fig:GC}.

\noindent
(3) Gravitational braking: There is a small probability that two single BHs
come to such a close distance that GW radiation dissipates a significant amount
of the orbital energy, leaving the two BHs gravitationally bound
\cite{turner77,quinlan87,kocsis06,oleary09,lee10,hong15}.  For GCs, and using
optimistic assumptions, these binaries contribute an event rate of
$0.06-20~{\rm Gpc^{-3}~yr^{-1}}$ in the LIGO band \cite{lee10,antonini16BS},
while in NSCs it has been estimated to range between $0.005-0.02~{\rm
Gpc^{-3}~yr^{-1}}$ \cite{tsang13}. The boundaries in Figure~\ref{fig:GC} for
BHBs formed via this mechanism can be calculated using the formulae of
\cite{oleary09}.  For that, we choose an initial relative velocity $v$ in the
range $\sigma_*<v<3\sigma_*$ and an initial impact parameter $b$ in the range
$0.3b_{\rm max}<b<0.99b_{\rm max}$ to account for the majority of the
encounters, because the encounter probability is proportional to $b^2$, and
$b_{\rm max}$ is the maximum impact parameter that leads to a bound binary.
The first LIGO detections, had they been originated via this mechanism, should
originate from the red area above the green line.

\begin{figure}
\resizebox{\hsize}{!}
          {\includegraphics[scale=1,clip]{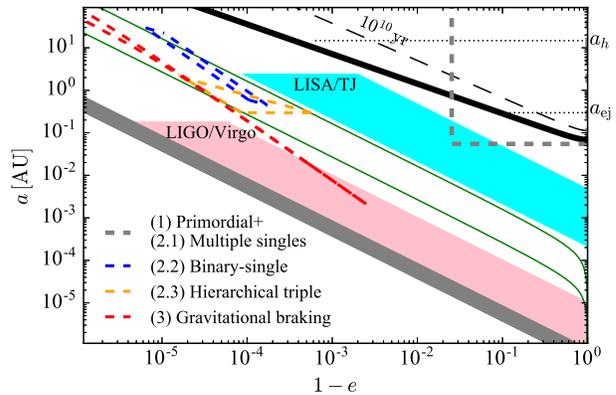}}
\caption
   {
Same as Fig.~\ref{fig:GC}, but for a BHB with masses as in GW151226 \cite{ligo16b},
i.e. $M_1=14~M_\odot$ and $M_2=8~M_\odot$. 
\label{fig:GC2}}
\end{figure}

{\it Discussions and conclusions.--}A joint detection of BHBs with LIGO/Virgo
and LISA/TJ would be desirable because of the science payback. In this paper we
show that the actual number of BHBs to be coincidentally detected is very
uncertain.  As Figure~\ref{fig:detectors} shows, LISA/TJ is already deaf to
mildly eccentric BHBs: For example, a BHB at milli-Hertz orbital frequencies
starting at $a \sim 10^{-3}$ AU and $0.7\lesssim e\lesssim0.9$ will also be
missed by LISA/TJ, but later be detectable by LIGO/Virgo.

BHBs can form via the five mechanisms which we discussed in the list of
possible formations. This allows us to pinpoint the regions in phase-space
which produce BHBs that eventually will merge via gravitational radiation. The
total area of these five regions is a small subset of phase-space.  It is an
error to assume that all binaries born in this subset are jointly detectable by
LIGO/Virgo and LISA/TJ.

Only a subset of that subset of phase-space will lead to successful joint
detections. This sub-subset depends on the masses of the BHBs.  We can see this
in Figures~\ref{fig:GC} and \ref{fig:GC2}. While in the first figure the
hierarchical triple gets into the LISA/TJ band, it does not in the second one.

On the other hand, up to 95\% of primordial and dynamical binaries (1 and 2.1
in the list of possible formations) are produced in the box delimited by grey
dashed lines. In that box, and in principle, the BHBs can lead to sources
jointly detectable by LIGO/Virgo and LISA/TJ. However, exceptions might occur
if a scatter results in a BHB jumping towards high eccentricities. This
probability has not been fully addressed. It requires dedicated numerical
scattering experiments with relativistic corrections (e.g.  \cite{ama10}), as
well as a proper star-cluster model to screen out BHBs that can decouple from
the stellar dynamics (e.g. our model as presented in Figures~\ref{fig:GC} and
\ref{fig:GC2}).

We have shown that mergers in GCs produced by the mechanisms (2.2), (2.3), and
(3) are inaudible to LISA.  The event rates corresponding to these mergers have
been largely discussed in the literature, but are uncertain, due to
questionable parameters, such as the cosmic density of GCs and the number of
BHs in them.  Nevertheless, it has been estimated that the rate could be as
large as $20~{\rm Gpc^{-3}~yr^{-1}}$ \cite{lee10}, while the current LIGO
detections infer a total event rate of $9-240~{\rm Gpc^{-3}~yr^{-1}}$.
Moreover, these mergers could also originate in NSCs,
\citep{kocsis06,MillerLauburg09,oleary09,tsang13,hong15,antonini12,addison15,antonini16},
and the event rates there are higher, up to $10^2~{\rm Gpc^{-3}~yr^{-1}}$
\citep{VL16}. 

Therefore, future multi-band GW astronomy should prepare for LIGO/Virgo BHBs
that do not have LISA/TJ counterparts. A non-detection by LISA/TJ is also
useful in constraining astrophysics: It puts a lower limit on the
eccentricities of the LIGO/Virgo sources, which according to
Figures~\ref{fig:GC} and \ref{fig:GC2} is about $0.005$. 

A deci-Hz detector, by covering the gap in frequencies between LISA/TJ and
LIGO/Virgo,  would drastically enhance the number of jointly detectable
binaries.

{\it Acknowledgement.}--This work is supported partly by the Strategic Priority
Research Program ``Multi-wavelength gravitational wave universe'' of the
Chinese Academy of Sciences (No. XDB23040100) and by the CAS President's
International Fellowship Initiative. PAS acknowledges support from the
Ram{\'o}n y Cajal Programme of the Ministry of Economy, Industry and
Competitiveness of Spain.  We thank Bence Kocsis and Fukun Liu for many
fruitful discussions, and Eric Peng for a thorough reading of our manuscript. 

\end{document}